\documentclass[12pt]{article}
\usepackage{latexsym,amsmath,amssymb,epsfig,graphicx}
\newcommand{\be}{\begin{equation}}
\newcommand{\ee}{\end{equation}}
\newcommand{\bea}{\begin{eqnarray}}
\newcommand{\eea}{\end{eqnarray}}
\newcommand{\kma}{\; ,}
\newcommand{\pkt}{\; .}

\newcommand{\calm}{{\cal M}}
 
\newcommand{\calg}{{\cal G}}

\newcommand{\bfx}{{\bf x}}

\newcommand{\eqn}[1]{(\ref{#1})}

\begin{document}


\begin{titlepage}
\begin{flushright}
hep-th/yymmnn \\
February 2011
\end{flushright}
\vspace{15mm}
\begin{center}
{\Large \bf
One-loop corrections to the Nielsen-Olesen vortex: finite length.
}
\\\vspace{8mm}
{\large  J\"urgen Baacke\footnote{e-mail:~
juergen.baacke@tu-dortmund.de}} \\
{  Fakult\"at Physik, Technische Universit\"at Dortmund \\
D - 44221 Dortmund, Germany
}\\
\vspace{10mm}
\bf{Abstract}
\end{center}
We consider the one-loop quantum corrections to the Nielsen-Olesen
flux tube of finite length $L$, by imposing periodic boundary conditions.
The calculations are based on a recent evaluation of these
quantum corrections to the string tension of an infinite vortex.
The finite length corrections are finite from the outset.
If the computation is restricted to the zero modes we obtain
the standard L\"uscher term $\pi/3L$ for a closed string. The
inclusion of the other fluctuation modes of Higgs and gauge fields,
using the numerically computed trace of the Euclidian Green's function,
 leads to corrections that decrease exponentially with $L$. 
We present numerical results for these corrections, discuss their
possible relevance, and the limitations of the approach.

\end{titlepage}


\section{Introduction}
\setcounter{equation}{0}
\label{sec:intro}
The vortex solution of the Abelian Higgs model in $(3+1)$ dimensions,
known from superconductivity \cite{Abrikosov:1957},
has been introduced in particle physics by 
Nielsen and Olesen \cite{Nielsen:1973cs}
as a possible model for strings.
Indeed the authors show that the vortex can be related to the
the bosonic Nambu-Goto string (see, e.g. \cite{Green:1987sp}).
This connection was mainly discussed on the classical
level, the corrections due to quantum fluctuations
were, there and later on, mostly considered already within string
theory. Within the underlying quantum field theory the one-loop
quantum corrections have been computed only recently
\cite{Baacke:2008sq}. The fact that the collective oscillations
of the string are related to zero modes of the quantum fluctuations
was already used qualitatively in Ref. \cite{Nielsen:1973cs}.
A detailed formulation of this connection has recently been presented
in Ref. \cite{Baacke:2010mj}; a further aspect has been
addressed there: in quantum field theory the renormalization of the energy
of collective fluctuations becomes part of the ordinary
renormalization programme, renormalization requires 
the inclusion of  the zero modes into the computation of
the one-loop corrections to the string tension, they are not
regularized and renormalized separately. 

While for this reason the contribution of the zero modes to the string tension
of an infinitely long string cannot be quantified separately
in quantum field theory, this
is no longer so for the finite corrections which appear
if one considers a string of finite length. It is the aim
of this work to elucidate this aspect and to determine
these corrections numerically. A vortex of finite
length can either be constructed as an open string or a closed
string. For an open string we would need to provide end caps, e.g., 
in the form of magnetic charges. It is hard to
imagine how one could possibly compute quantum corrections
to such a configuration. The same would hold for a
closed string in the form of a torus. The technique developed
in Ref. \cite{Baacke:2008sq} can be applied, however,
to a vortex of finite length with periodic boundary conditions.
This can be considered as an approximation to a realistic closed string 
if the length of the string is much larger than its transverse
extension. Such a computation will be the main subject of this
article.

The text is organized as follows:
In Sec. \ref{sec:model} we recall some basic formulae of the model and its 
quantum fluctuations, referring mainly to Ref. \cite{Baacke:2008sq} for
all details; in Sec. \ref{sec:basics} we develop the formalism
for computing the quantum corrections to the energy of a finite string,
i.e., those one-loop corrections that are not already included
in the one-loop corrections to the string tension;
explicit calculations, analytical and numerical, are
presented in Sec. \ref{sec:numericalandanalytical};
the results are discussed in Sec. \ref{sec:discussion};
we conclude with a brief summary in Sec. \ref{sec:summary}.

 
\section{The model}
\setcounter{equation}{0}\label{sec:model}
The Abelian Higgs model in (3+1) dimensions is
defined by the Lagrange density 
\begin{equation}
  {\cal L}=-\frac{1}{4}F_{\mu\nu}F^{\mu\nu}
+\frac{1}{2}(D_\mu\phi)^*D^\mu\phi-\frac{\lambda}
{4}\left(|\phi|^2-v^2\right)^2\; .  
\end{equation}
Here $\phi$ is the complex scalar Higgs field and
\bea
F_{\mu\nu}&=&\partial_\mu A_\nu-\partial_\nu A_\mu\kma \\
D_\mu&=&\partial_\mu-igA_\mu 
\pkt\eea 
The particle spectrum consists of Higgs bosons of mass
$m_H^2=2\lambda v^2$ and vector bosons of mass $m_W^2=g^2v^2$.
We denote the ration of these masses as $\xi=m_H/M_W=2\lambda/g^2$.
The  vortex solution \cite{Abrikosov:1957,Nielsen:1973cs}
is defined, in the singular gauge, by the cylindrically symmetric
ansatz 
\footnote{ We use Euclidean notation for the transverse components, so
$A^\perp_1\equiv A^1=-A_1$ etc. } 
\begin{eqnarray}
A^{{\rm cl}\perp}_i (x,y,z)&=&\frac{\varepsilon_{ij}x^\perp_j} 
{gr^2}\left[A(r)+1\right] \;\;\; i=1,2 \\
\phi^{\rm cl}(x,y,z)&=&vf(r) \; .
\end{eqnarray}
where $r=\sqrt{x^2+y^2}$ and $\varphi$ is the polar angle. 
The equations of motion for $f(r)$ and $A(r)$ can be solved
numerically, see, e.g., Ref. \cite{Baacke:1994bk}. In terms of these 
functions the classical string tension takes the form
\begin{eqnarray}
\sigma_{\rm cl}&=&\pi v^2
 \int^{\infty}_{0}\!\!\! dr \left\{\frac{1}{rm_W^2}\left[
\frac{dA(r)}{dr}\right]^2\!\!+r\left[\frac{df(r)}{dr}\right]^2\!\!
+\frac{f^2(r)}{r}\left[A(r)+1\right]^2\!\! \right. \nonumber \\ 
&+&\left. \frac{rm^2_H}{4}\left[
f^2(r)-1\right]^2\!\right\}\pkt
\label{eq:classtens} \end{eqnarray}
The parameter dependence can be written in the form 
$\sigma_{\rm cl}=(\pi/g^2)h(\xi)$, where $h(\xi)$ varies from
$0.75$ at $\xi=0.5$ to $1.34$ for $\xi = 2$. Here and elsewhere we use
units such that $m_W=1$.

The fluctuations around this classical solution consist of those
of the real and imaginary part of the Higgs field and those of the
transversal, longitudinal and timelike components of the gauge field;
furthermore the gauge fixing introduces the corresponding Faddeev-Popov
fields. In the fluctuation energy the Faddeev-Popov contributions
cancel those of longitudinal and timelike components of the gauge field,
so that these are irrelevant here \cite{Baacke:2008sq}.
The remaining fluctuations form a $4\times 4$ coupled system. 
As the classical solution is independent of time and $z$ 
the Euclidian fluctuation operator has the general form
\be 
\calm_{ij} = - (\partial_\tau^2 +\partial_z^2)\delta_{ij} + \calm_{\perp i j}
\pkt
\ee
The transversal fluctuation operator $\calm_{ \perp ij}$
is identical to the one  of the instanton in the $2$-dimensional model;
is has been presented in detail in  Refs. \cite{Baacke:1994bk,Baacke:2008sq}.
The Green' s function of $\calm_{ij}$ has, in momentum space,
the formal representation
\be \label{eq:greenformal}
\calg_{ij}(\bfx_\perp,\bfx'_\perp,k,\nu)=
\sum_\alpha \frac{\psi^\alpha_i(\bfx_\perp)\psi_j^{\alpha}
{}^\dagger(\bfx_\perp')}{\nu^2+k^2+\lambda_\alpha^2}
\ee
where $\lambda_\alpha^2$ and $\psi^\alpha_i(\bfx_\perp)$ 
are the eigenvalues and eigenfunctions of $\calm_{\perp ij}$,
respecively. The trace of this Green' s function has been computed 
numerically in Ref. \cite{Baacke:2008sq}, not via \eqn{eq:greenformal},
but using a Jost function formalism adapted to coupled systems, 
see Ref. \cite{Baacke:1990zu} and the  Appendix  of Ref.\cite{Baacke:1995hw}.


\section{The energy of the vortex of finite length: basic relations}
\setcounter{equation}{0}
\label{sec:basics}
We now establish the formalism for computing the 
quantum corrections to the energy of a vortex of finite length.
As announced in Sec. \ref{sec:intro}, we do this in an approximate 
way by imposing periodic boundary conditions in the longitudinal 
coordinate $z$.

Starting point is the formal definition
\be
E_{\rm fl}(L)=\sum_{n=-\infty}^\infty\sum_\alpha
\frac{1}{2}\left(E_\alpha(k_n) -E_{0\alpha}(k_n)\right)
\kma\ee
where $E_\alpha$ and $E_{0\alpha}$
are the energies of the eigenmodes of the
fluctuation operators around the vortex and in the vacuum, 
respectively. The variable $k_n$ is the momentum in the
longitudinal direction of the vortex, which for periodic boundary
takes the values $k_n=2\pi/L$.
The energies of the eigenmodes have the form
\be 
E_\alpha(k_n)=\sqrt{k_n^2+\lambda_\alpha^2}
\ee
and analogously for $E_{0\alpha}(k_n)$. Here $\lambda_\alpha^2$ are
the eigenvalues of the fluctuation operator
$\calm_\perp$ in the transverse variables, as defined in Sec.
\ref{sec:model}.

We have introduced in the previous section the
Green' s function of the fluctuation operator and its
formal representation \eqn{eq:greenformal}.
We define a function $F(k_n,\nu)$ as
\begin{equation}
\label{fk3nudef}
F(k_n,\nu) = \int d^2x_\perp~ 
{\rm Tr}~ ({\cal G}(\bfx_\perp,\bfx_\perp,k_n,\nu)-
{\cal G}^0(\bfx_\perp,\bfx_\perp,k_n,\nu))
\kma\end{equation}
Here $\calg_0$ is the free Green' s function. We denote the
eigenvalues of the free fluctuation operator by 
$(\lambda_\alpha^{(0)})^2$. We then obtain the relation
\begin{equation}
-\int_{-\infty}^{\infty}\frac{d\nu \nu^2}{2\pi} F(k_n,\nu) =
\sum_\alpha\frac{1}{2}\left[\sqrt{k_n^2+\lambda_\alpha^2}
-\sqrt{k_n^2+(\lambda^{(0)}_\alpha)^2}\right]
\kma \end{equation}
which is the basis of our numerical computation.
In terms of $F(k_n,\nu)$  we find for the energy of a 
string of finite length
\be
E_{\rm fl}(L)=-\int_{-\infty}^\infty\frac{d\nu}{2\pi}\nu^2
\sum_{n=-\infty}^\infty F(2\pi n/L,\nu)
\pkt\ee 
Actually $F(k_n,\nu)$ only depends on $k_n^2+\nu^2$. So with the 
definitions $p=\sqrt{k_n^2+\nu^2}$ and $F(p)\equiv F(k_n,\nu)$
we may write this as
\be
E_{\rm fl}(L)=-\int_{-\infty}^\infty\frac{d\nu}{2\pi}\nu^2
\sum_{n=-\infty}^\infty F(p_n)
\kma \ee
with $p_n=\sqrt{\nu^2+(2\pi n/L)^2}$.
This can easily be converted into a weighted integral over $F(p)$
via
\be\label{eq:weightint}
E_{\rm fl}(L)=-\int_0^\infty\frac{p ~dp}{\pi}
w(p) F(p)
\kma\ee
with
\be
w(L,p)=\sum_{n=-\infty}^\infty\sqrt{p^2-(2\pi n/L)^2}
\Theta(p^2-(2\pi n/L)^2)\pkt\ee

\begin{figure}[htb]
\vspace{12mm}
\begin{center}
\includegraphics[scale=0.4]{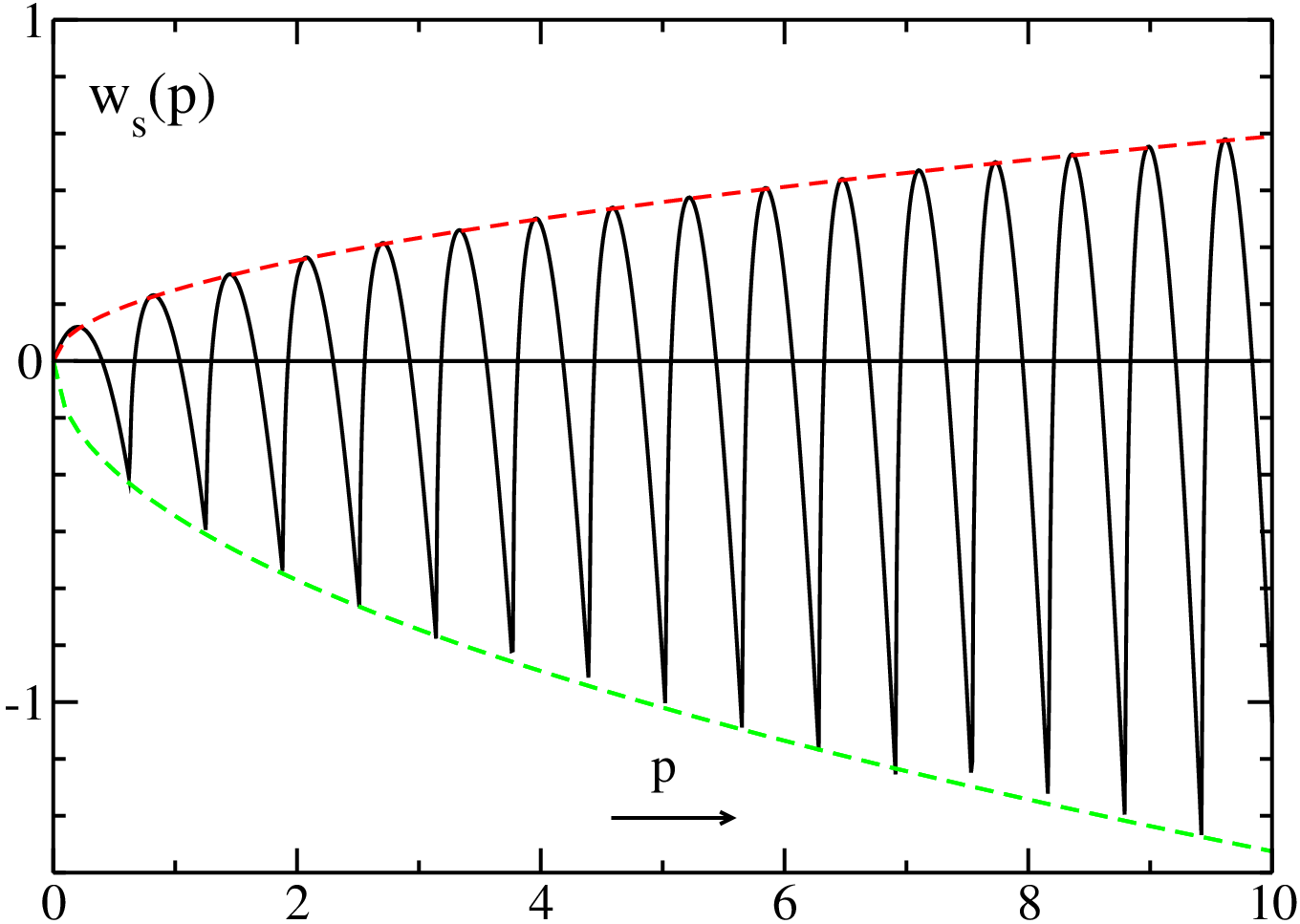}
\end{center}
\hspace{3mm}
\caption{\label{fig:wsubofp} 
The subtracted weight function $w_{\rm s}(p)$.
Solid line: $w_{\rm s}(p)$. The dashed lines indicate
a square root behaviour of the maxima and minima.}
\end{figure}

The function $F(p)$ has been computed numerically in Ref. \cite{Baacke:2008sq}
for various values of the parameter $\xi=m_H/m_W$. At small $p$ it behaves as
$2/p^2$, due to the presence of the two translation zero modes.
At large $p$ it behaves as $a/p^2$, where the coefficient $a$ is determined
by the leading-order Feynman diagrams, see Eq. (7.1) of Ref.
 \cite{Baacke:2008sq}.
One easily convinces oneself that the weighted integral
in Eq. \eqn{eq:weightint} is quadratically divergent. In the limit 
$L\to \infty$ the sum over $n$ can be replaced by an integral and
one finds
\be
\lim_{L\to \infty} \frac{w(L,p)}{L}=\frac{1}{2\pi}
\int_{-\infty}^\infty dx\sqrt{p^2-x^2}\Theta(p^2-x^2)=\frac{1}{2\pi}
\frac{\pi}{2}p^2
\pkt\ee
This limit yields that part of the string energy which is 
proportional to its length, $E \propto L \sigma$ and defines
the string tension, or energy per length, 
of the vortex of infinite length \footnote{This
expression for $\sigma_{\rm fl}$ is of course divergent. 
$F(p)$ has to be used in subtracted form, and the divergent
parts have to be renormalized, this is the subject of Ref.
\cite{Baacke:2008sq}.}

\be
\sigma_{\rm fl}=-\int_0^\infty \frac{p^3 ~dp}{4\pi}F(p)
\pkt \ee
Here we want to determine the corrections
which arise for finite length. It is convenient, therefore,
to subtract from $E_{\rm fl}(L)$ 
the term  $\sigma_{\rm fl} L$ . This can be done by 
redefining the weigth $ w(L,p)$ by subtracting
 the asymptotic weight:
\be 
 w_{\rm s}(L,p)= w(L,p)-\frac{L}{4}p^2
= \sum_{n=-\infty}^\infty\sqrt{p^2-(2\pi n/L)^2}
\Theta(p^2-(2\pi n/L)^2)-
\frac{L}{4} p^2\pkt\ee
The finite length correction to the energy
then becomes
\be \label{eq:weightintsub}
\Delta E_{\rm fl}(L)= E_{\rm fl}(L)-\sigma_{\rm fl}L=
-\int_0^\infty\frac{p ~dp}{\pi}
w_{\rm s}(L,p) F(p)\pkt
\ee
In this way we have gotten rid of the divergences.
Indeed we have already used all counter terms of quantum
field theory in order to obtain a finite string tension
$\sigma_{\rm fl}$, as described in detail in Ref. \cite{Baacke:2008sq}.
One can verify explicitly, as we will see in the next section,
 that for a function $F(p)$ which
asymptotically behaves as $c/p^2$ the weighted integral
of Eq. \eqn{eq:weightintsub} is UV finite. The function $w_{\rm s}(p)$
does not tend to zero at large $p$, it oscillates with a
period of $\Delta p \simeq 2\pi/L$, without being strictly periodic.
The oscillations are due to the fact that more and more terms are included
into the sum over $n$. $ w_{\rm s}(p)$  is displayed in 
Fig. \ref{fig:wsubofp}, for $L=10$.

\section{Numerical and analytical calculations}
\label{sec:numericalandanalytical}
\setcounter{equation}{0}
As the function $F(p)$ is know numerically from previous
computations it seems a straightforward matter to evaluate
the weighted integral of Eq. \eqn{eq:weightintsub}. However,
this integral is subtle numerically, due to the oscillating 
and spiky weight function. It is useful, therefore, to begin
with some related analytical and numerical calculations.

\begin{figure}[htb]
\vspace{12mm}
\begin{center}
\includegraphics[scale=0.4]{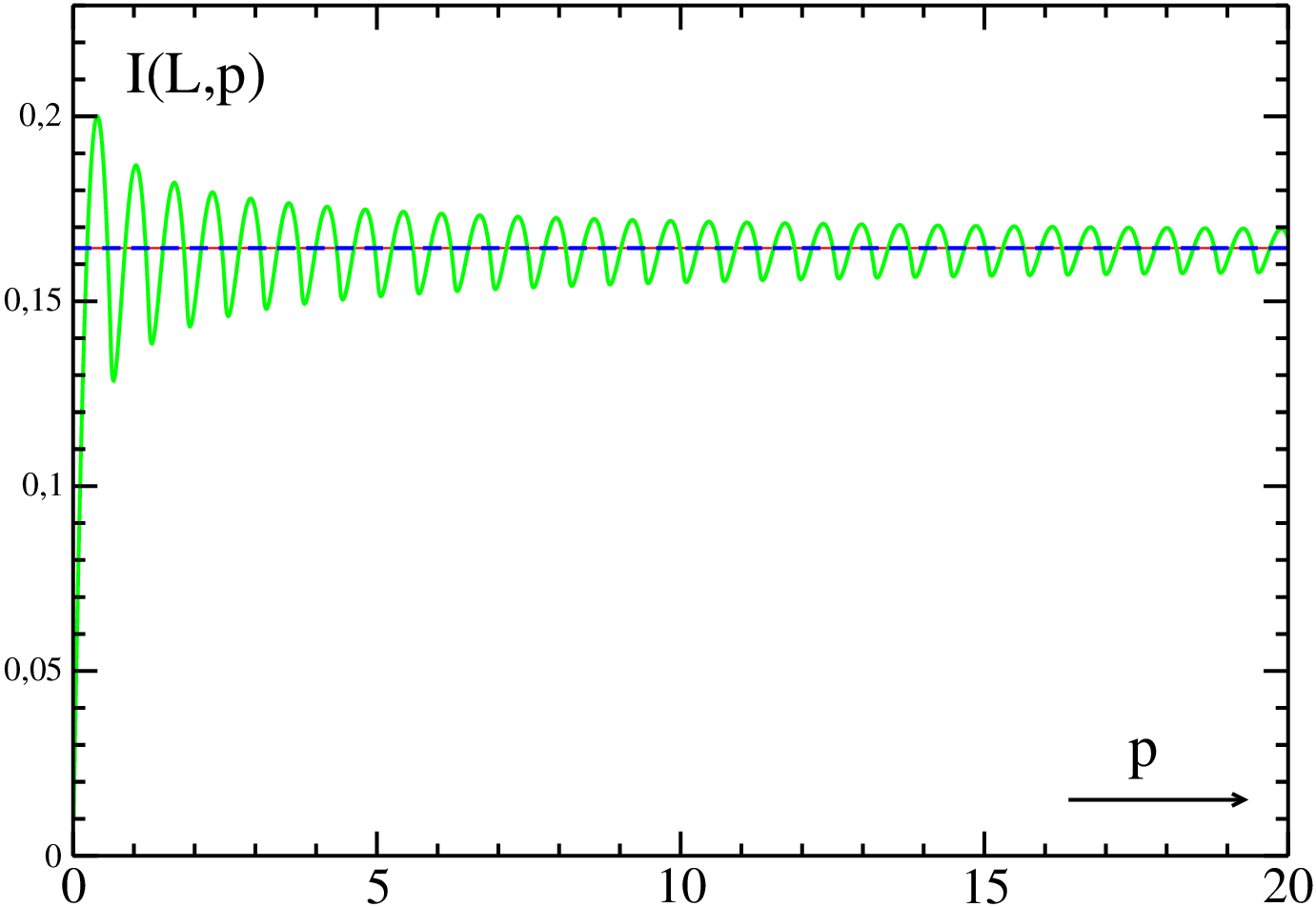}
\end{center}
\hspace{3mm}
\caption{\label{fig:ioflp} 
The integral $I_0(L,p)$ of Eq. \eqn{eq:Idef}.
Solid line: the function $I_0(L,p)$; dashed horizontal line: the
asymptotic limit $0.1644$.}
\end{figure}

The function $F(p)$ is dominated, at small $p$ by the two
zero modes describing collective oscillation, and it is instructive
to compute their contribution to $\Delta E_{\rm fl}(L)$.
For $\sigma_{\rm fl}$ such a separate computation was not possible,
as the contribution of the collective oscillations is infinite.
When computing the correction $\Delta E_{\rm fl}$ this problem does not arise.
We define the integral
\bea\nonumber
&&I_0(L,p) =
\int_0^p\frac{p ~dp}{p^2}
 w_{\rm s}(L,p) 
\\\nonumber
&&=2 \sum_{n=1}^{[Lp/2\pi]}
\left[\sqrt{p^2-(2\pi n/L)^2}-\frac{2\pi n}{L}
\arccos\frac{2\pi n}{Lp}\right]
\\\label{eq:Idef}
&&+p -\frac{\pi}{4}\frac{L}{2\pi}p^2
\eea
and its limit as $p\to \infty$ 
\be
\bar I_0(L)=\lim_{p\to \infty}I_0(L,p)
\pkt\ee
In terms of $\bar I_0(L)$ the contribution of the two zero modes is
\be \label{eq:DeltaEI}
\Delta E_{\rm coll}(L)=-\frac{2}{\pi}\bar I_0(L)
\pkt \ee
 At finite $p$ the sum on the right hand side of Eq. \eqn{eq:Idef} extends
up to $N=[Lp/2\pi]$.  It can be evaluated using the Euler-Maclaurin 
summation formula.
One finds
\be \label{eq:limioflp}
\bar I_0(L)=\frac{\pi^2}{6 L}
\pkt\ee
We display the integral $I_0(L,p)$ in Fig. \ref{fig:ioflp} for $L=10$.
One sees that the integral oscillates as a function of $p$, the width
of these oscillations narrows only slowly, as $1/\sqrt{p}$, and the 
band is slightly asymmetric.
At $p=500$ the width is of order $0.002$ and the mean value
can be read off, by taking the average of maxima and minima,
 with a precision of order $0.0001$. For $L=10$ one finds
$I_0(L,p)\simeq 0.1644$; this is consistent to four digits with the
analytic result $\pi^2/6L=\pi^2/60=0.164493$. The same value
is found by evaluating the weighted integral via numerical
integration. 

Using Eq. \eqn{eq:DeltaEI} we find
\be\Delta E_{\rm coll}(L)=-\pi/3L
\pkt
\ee 
This is the L\"uscher term \cite{Luscher:1980fr,Luscher:1980ac} 
for a closed string (see, e.g., \cite{Athenodorou:2010cs}). 
For a closed string the mode energies are $2\pi n/L$ instead of
$\pi/L$ for an open string, and the oscillating modes can propagate
in two directions, up and down the $z$ axis. This explains the
factor $4$ relative to the usually quoted value of $\pi/12 L$.
We note that our result is obtained by evaluating an expression 
that is finite from the outset. This is analogous to finite temperature 
corrections which likewise do not need a regularization and
renormalization.

An important property of our weighted integral $I(L,p)$ is
apparent in Fig. \ref{fig:ioflp}:
the mean value between maxima and minima of the oscillations
attains the final value of the integral already at very low values of $p$
, i.e., after a few oscillations with period $2 \pi/L$.
This means in general, that in a region where  $F(p)\propto 1/p^2$ the
integral of Eq. \eqn{eq:weightintsub} will just oscillate around an
almost constant average value. As $F(p) \simeq c/p^2$ for 
large $p$ this means that in the asymptotic regime
the average value of the integral reaches its asymptotic limit quickly. 

\begin{figure}[htb]
\vspace{12mm}
\begin{center}
\includegraphics[scale=0.40]{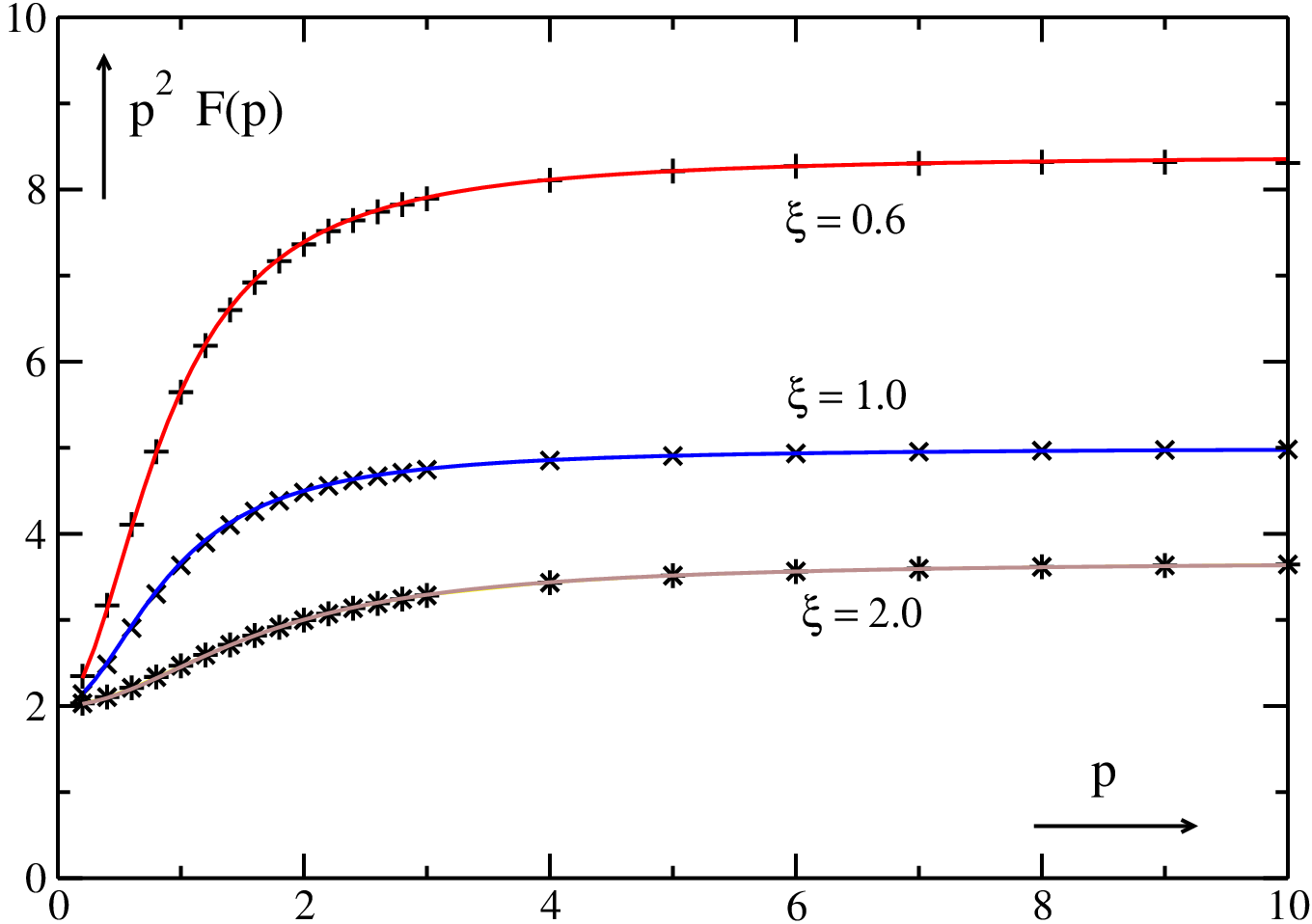}
\end{center}
\hspace{3mm}
\caption{\label{fig:fits} 
The fits to the function $ F(p)$ of Eq. \eqn{eq:polefit}.
We display the data and the fits for $p^2 F(p)$
for $\xi=0.6\kma\xi=1.0$ and $\xi=2.0$. Symbols
($+, \times, *$): data; solid
lines: fits according to Eq. \eqn{eq:polefit}}
\end{figure}

In order to perform the weighted integral for the
realistic case of Eq. \eqn{eq:weightintsub}
we need $F(p)$ for a very narrow grid of values $p$,
$\Delta p \ll 2\pi/L$, as  the weight varies strongly and the integration
implies subtle cancellations. The numerical
computation or Ref. \cite{Baacke:2008sq} has provided values only on
a relatively coarse grid of values $p$. As a computation of $F(p)$ requires 
substantial CPU time, a few minutes for each value of $p$, 
while the function itself is varying smoothly, it is convenient
to use fits through the existing data points. One may use
spline fits, but it turns out that a simple
parameterization 
\be \label{eq:polefit}
F(p)=2/p^2+d/(p^2+\lambda^2) 
\ee
gives a surprisingly good global fit to the data, for
all values of $\xi=m_H/m_W$.
The ansatz can be understood as a fit where the effect of all
higher modes, possible bound states and continuum, is simulated 
by just one pole with an effective degeneracy $d$ at an energy $\lambda$ on 
the imaginary $p$ axis, which contains the physical cut.
The symbol $\lambda$ refers to Eq. \eqn{eq:greenformal}, 
$\lambda^2$ being an eigenvalue of the transversal fluctuation operator.
The parameters $d$ and $\lambda^2$ 

for various values of $\xi$, determined by a fit-by-eye,
are given in Table \ref{table:fitparams}. The fits and 
the numerical data are displayed in Fig. \ref{fig:fits}
for three values of $\xi=m_H/m_W$. If necessary the approach
could be improved systematically by including more poles 
into the ansatz \eqn{eq:polefit}.
 This type of interpolation can be considered as a
 Pad\'e approximation (see, e.g., Ref. \cite{Press}, Sec. 5.12)
and is well adapted for functions with a cut in the complex plane.
The parameters could be determined, e.g.,  by least squares methods. 

\begin{figure}[htb]
\vspace{12mm}
\begin{center}
\includegraphics[scale=0.40]{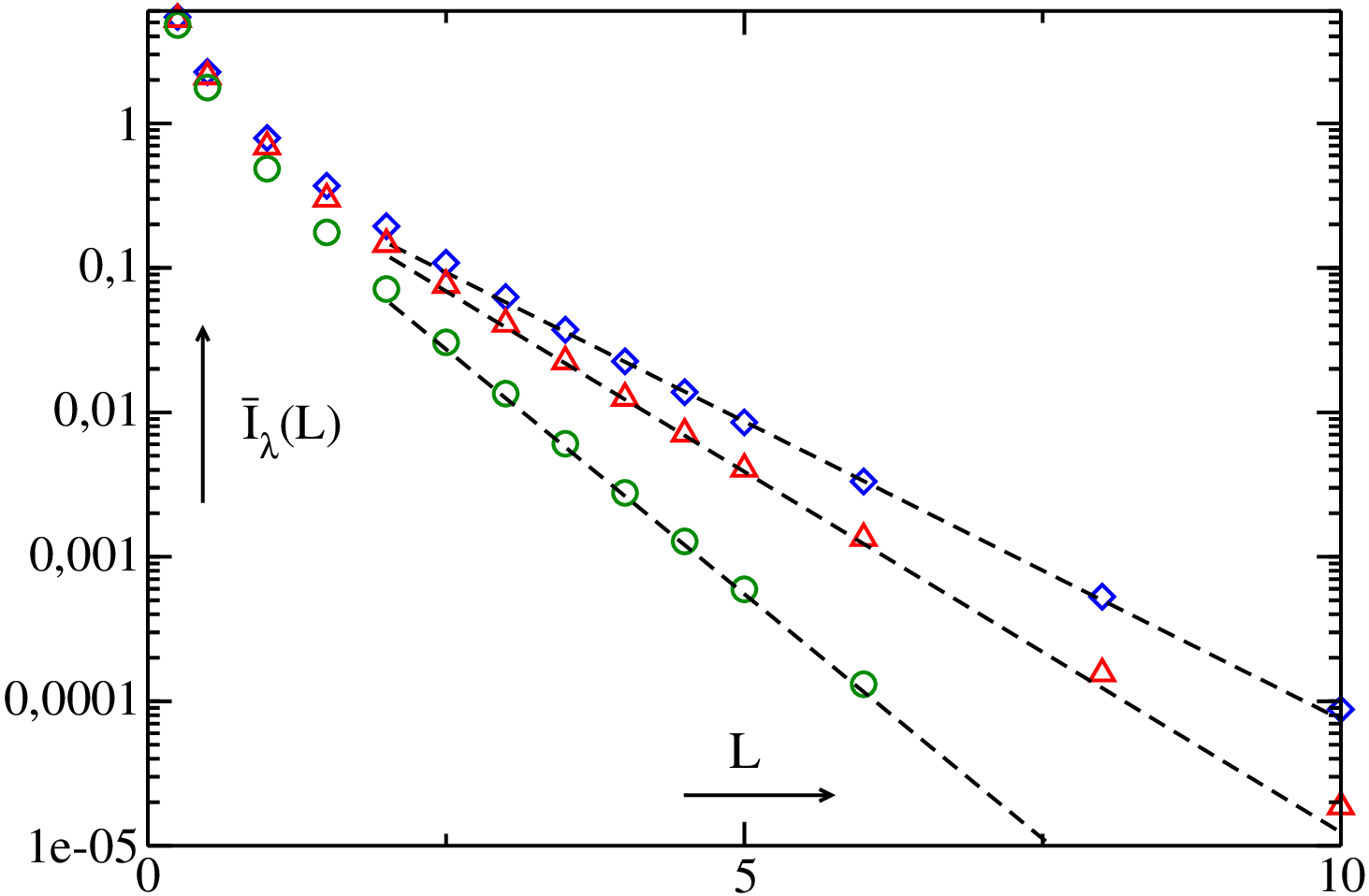}
\end{center}
\hspace{3mm}
\caption{\label{fig:IlamofL} 
The integral $\bar I_\lambda(L)$.
diamonds: $\lambda^2=0.7$; triangles: $\lambda^2=1.0$; circles:
$\lambda^2=2.0$; the dashed lines indicate a simple exponential
behaviour, as specified in the text. }

\end{figure}

A considerable advantage of our fit is the fact that we can  do
the integral over $p$ for the second pole term analytically, as well.
We thus can avoid a very subtle numerical integration. One finds
\bea\nonumber
&&I_\lambda(L,p)\equiv 
\int_0^p\frac{p ~dp}{p^2+\lambda^2}
 w_{\rm s}(L,p) 
\\\nonumber
&&=2 \sum_{n=1}^{[Lp/2\pi]}
\left[\sqrt{p^2-(2\pi n/L)^2}-
\sqrt{(2\pi n/L)^2+\lambda^2}
\arccos\frac{\sqrt{(2\pi n/L)^2+\lambda^2}}{p}\right]
\\\label{eq:Ilamdef}
&&+p -\lambda \arccos\frac{\lambda}{p}
-\frac{L}{8}\left(p^2-\lambda^2\ln\frac{p^2+\lambda^2}{\lambda^2}\right)
\pkt
\eea
The delicate cancellations between the sum and the other terms on the
right hand side can be avoided
by subtracting $I_0(L,p)$ and adding its asymptotic limit $\pi^2/6L$
of Eq. \eqn{eq:limioflp}. Of course the resulting expression
\bea\nonumber
&& \tilde I_\lambda(L,p)=I_\lambda(L,p)-I_0(L,p)+\frac{\pi^2}{6L}
\\\nonumber
&&=2 \sum_{n=1}^{[Lp/2\pi]}
\left[\frac{2\pi n}{L}\arccos\frac{2\pi n}{Lp}-
\sqrt{(2\pi n/L)^2+\lambda^2}
\arccos\frac{\sqrt{(2\pi n/L)^2+\lambda^2}}{p}\right]
\\\label{eq:Ilamdef}
&& -\lambda \arccos\frac{\lambda}{p}
+\frac{L}{8}\lambda^2\ln\frac{p^2+\lambda^2}{\lambda^2}+\frac{\pi^2}{6L}
\eea
is not identical to $I_\lambda(L,p)$, but it has the same limit
as $p\to \infty$. 
The oscillations present in $I_\lambda(p,L)$ are considerably suppressed
in $\tilde I_\lambda(p,L)$ and the limit $p\to\infty$, which we denote by
$\bar  I_\lambda(L)$, can be evaluated numerically without problems. 
To obtain the corrections to the
L\"uscher term in $\Delta E_{\rm fl}(L)$, we have to multiply the result
by $-d/\pi$, so that, with our fit to $F(p)$ we get
\be
\Delta E_{\rm fl}(L)=-\frac{\pi}{3L}-\frac{d}{\pi}\bar I_\lambda(L)
\pkt\ee
In Fig. \ref{fig:IlamofL} we plot $\bar I_\lambda (L)$  as a function of
$L$ for $\lambda^2= 0.7, ~1.0$ and $2.0$, values which are in the range
of the fit parameters given in Table \ref{table:fitparams}.
The results obtained display a roughly exponential decrease with $L$.
This is not unexpected; had we imposed periodicity in time
instead of periodicity in $z$ we would have expected
 the thermal corrections to display such an exponential behaviour. 
We have to take into
account that we do not consider the contribution of
a single energy level, but the contribution of a cut
generated by the pole in the transversal Green' s function.   
So the correction is not expected to be described by a
simple exponential function $a\exp(-\alpha L)$. However,
the ``effective'' logarithmic slope $\alpha$ is in the expected range 
of values. For the
range $2 \le L \le 6$ we have $\alpha\simeq 0.95$ for $\lambda^2=0.7$,
 $\alpha\simeq 1.15$ for $\lambda^2=1$ and 
 $\alpha\simeq 1.55$ for $\lambda^2=2$. The straight dashed lines
in Fig. \ref{fig:IlamofL} indicate this behaviour.
We would naively have expected $\alpha= \lambda= 0.837$, $1.0$ and $1.414$, 
respectively. Indeed for $L > 6$ the slopes $\alpha$ decrease and may 
attain these values as $L\to \infty$. For $L < 2$ on the other hand 
the effective slopes increase.

\section{Discussion of the results}
\begin{figure}[htb]
\vspace{12mm}
\begin{center}
\includegraphics[scale=0.40]{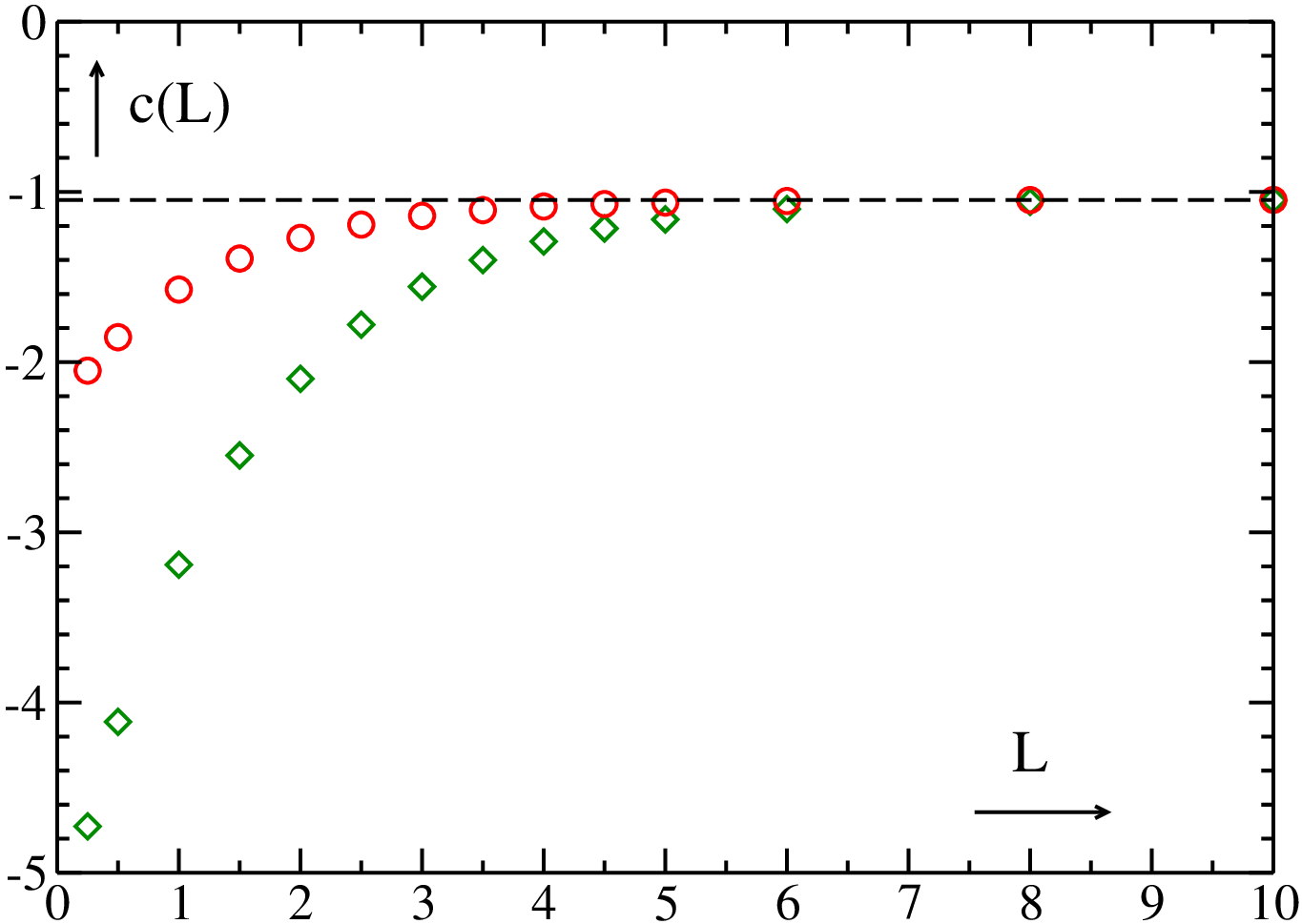}
\end{center}
\hspace{3mm}
\caption{\label{fig:ratio} 
The coefficient $c(L)$.
circles: $\xi=0.5$; diamonds: $\xi=1.2$; triangles: $\lambda^2=1.0$; circles:
$\lambda^2=2.0$; the dashed lines indicate a simple exponential
behaviour, see text. }
\end{figure}

\setcounter{equation}{0}
\label{sec:discussion}
Using the results of the previous section the final
results for $\Delta E_{\rm fl}(L)$ can be easily obtained, using
the parameters $d$ and $\lambda^2$ from Table \ref{table:fitparams}.
We present, in Fig. \ref{fig:ratio} the function 
$c(L)=L \Delta E_{\rm fl}(L)$
for $\xi=0.5$ and $\xi=1.2$. The function $c(L)$ can be considered
 as the coefficient of an ``effective'' L\"uscher term 
$c(L)/L$. With this formulation we follow the presentation of 
lattice measurements of this term in QCD in Ref. \cite{Luscher:2002qv}.
As one sees from table \ref{table:fitparams}
the values for $d$ and $\lambda$ are $8.5$ and $0.7$,
respectively, for $\xi=0.5$, for $\xi=1.2$ they
are $2.4$ and $1.0$, respectively. For the first parameter set
the corrections to a pure L\"uscher term are sizeable even at
$L=4$ and have decreased to the $10 \%$ level at $L=6$,
while for $\xi=1.2$ they have decreased to this level
already at $L=3$. For higher values of $\xi$ the effective
degeneracies become smaller and so does the deviation from
$c=-\pi/3$. 

There are various aspects under which we can consider the results
of these computations. It is satisfactory, at first, 
that the L\"uscher term appears in
a straightforward way. As this term has been measured for QCD strings
\cite{Luscher:2002qv} it is certainly a part of established physics.
 The fact that the corrections due to higher fluctuations appear here within
the same formalism and in an analogous way gives us confidence
that these terms as well are not artefacts of the approximation
but terms that would appear as a result of a suitable
measurement.

Of course our analysis is an approximation; we
see two essential limitations: 
(i) we have to require $L$
to be ``much'' larger than the transversal extension of the vortex
and (ii) the corrections have to be small enough for the
semiclassical approximation to be reliable.

The transversal extension of the vortex is of the order of
 $\max (1/m_H,1/m_W)$. In our computation the mass $m_W$ is set 
to unity, so the transversal extension is $R\simeq\max (1,1/\xi)$;
for the range of $\xi$ considered here this is between $0.5$ and $2$.
An optimistic guess of ``much larger'' could be $L>R$ and a 
pessimistic one $L > 10 R$. In the second case our correction will be
be negligible for all values of $\xi$. The range of validity also
depends on what is really measured. If, e.g.,  the corrections were 
measured on a lattice with the same periodic boundary conditions, 
then our analysis would be valid in the whole range of parameters
considered here. The limitation of the approximation depends,
unfortunately, mainly on the effects which we have neglected: the
influence of end caps for an open string, and the effects of
curvature for a closed vortex. Their magnitude will we difficult to 
estimate.

The second limitation is the validity of the semiclassical
approximation. The classical string tension is given by
$\pi/g^2$ times a number close to unity. The one-loop
corrections to the string tension were found, in Ref. \cite{Baacke:2008sq}
to be smaller than $0.5$, so this correction is small
even for $g$ as large as unity. The corrections for a
finite string, as found here, are much larger. Of course the ratio
depends on $L$: $\sigma_{\rm cl}$ is multiplied by $L$ while
$c(L)$ is divided by $L$. But for $g\simeq 1$ and $L\simeq 2$ the
ratio of these contributions to the string energy is not small.
In any case, for $L$ small enough the L\"uscher term and the even
larger corrections  will become comparable or exceed the classical
energy and the semiclassical approximation will break down.
For the lattice results it is found that the absolut
e value
of  $c(L)$ becomes {\em smaller} than $\pi/12$ at small $L$, 
and possibly tends to zero. In Ref. \cite{Luscher:2002qv} this 
behaviour is described by a relation derived using the QCD 
renormalization group and therefore relies on asymptotic freedom. 
Such an analysis does not apply here. The absolute value
of our coefficient $c(L)$ increases at small $L$ and it is hard to
see how this could be different as the parameters $d$ are positive
throughout. A higher order resummation seems out of scope.


\section{Summary}
\setcounter{equation}{0}
\label{sec:summary}
We have computed the one-loop corrections to the energy of
a Nielsen-Olesen vortex of finite length. More precisely: we have computed 
the corrections that are not already included in the string tension.
The latter were the subject of Ref. \cite{Baacke:2008sq}.
The corrections computed here are finite from the outset and, therefore,
do not depend on renormalization conditions.
The leading order correction at large $L$ is the L\"uscher term which here
takes the form $\pi/3L$ as appropriate for a closed string. 
The further corrections decrease exponentially at large $L$ but can be, 
depending on the parameters, relevant for small $L$ and intermediate $L$.
Within our computational framework they appear on the
same footing as the L\"uscher term and are related to the internal
structure of the vortex. 

We have discussed briefly the limitations of the approach and conclude
that, depending on the parameters of the model, there is a window
in $L$ where the corrections to the L\"uscher term are relevant and 
where their computation is reliable.

We would finally like to point out that the method used here can be 
applied in a similar way to other vortex configurations, like
cosmic strings (see \cite{Copeland:2009ga} for a recent review). 
In fact the only information specific to the Abelian Higgs model 
was contained in $F(p)$, the trace of the Green' s function
of the fluctuation operator. Furthermore, the behaviour of $F(p)$ 
at small $p$ is determined by the zero modes and its asymptotic 
behaviour can be obtained from leading order Feynman graphs. So
semi-quantitative estimates are easily accessible.  

\begin{table}
\begin{center}
\begin{tabular}{|r|r|r|}
\hline
$\xi$& $d$&$\lambda^2$
\\
\hline
$0.5$&8.5&0.7\\
$0.6$&6.4&0.75\\
$0.7$&5.0&0.75\\
$0.8$&4.1&0.8\\
$0.9$&3.48&0.8\\
$1.0$&3.0&0.8\\
$1.1$&2.65&0.9\\
$1.2$&2.4&1.0\\
$1.3$&2.22&1.1\\
$1.4$&2.06&1.2\\
$1.5$&1.94&1.3\\
$1.6$&1.87&1.6\\
$1.7$&1.8&1.85\\
$1.8$&1.75&2.1\\
$1.9$&1.71&2.35\\
$2.0$&1.68&2.7\\
\hline
\end{tabular}
\end{center}
\vspace*{3mm}
\caption{\label{table:fitparams} The parameters of a pole fit
to $F(p)$, Eq. \eqn{eq:polefit}}
\end{table}

\newpage

\bibliography{novfl}
\bibliographystyle{h-physrev4}

\end{document}